\documentclass[prl,twocolumn,superscriptaddress,reprint]{revtex4-1}
\usepackage{graphicx,amssymb,amsmath,epsf,bm,cprotect,comment}
\epsfclipon
\newcommand{\unit}[1]{~\mathrm{#1}}
\renewcommand{\(}{\left(}
\renewcommand{\)}{\right)}

\newcommand{\expct}[1]{\langle{#1}\rangle}

\newcommand{\cum}[1]{\langle{#1}\rangle_\mathrm{c}}

\newcommand{\diff}[2]{\frac{\mathrm{d} #1}{\mathrm{d} #2}}   
 
\newcommand{\prt}[2]{\frac{\partial{#1}}{\partial{#2}}}
\newcommand{\prts}[3]{\frac{\partial^{#3}#1}{\partial{#2}^{#3}}}
\newcommand{\tod}{\stackrel{d}{\to}}

\newcommand{\hlab}{{h_\text{lab}}}

\newcommand{\tlabinit}{{t_\text{lab}^\text{init}}}
\newcommand{\FBM}{FBM$_{1/3}$}

\renewcommand{\eqref}[1]{Eq.~(\ref{#1})}

\newcommand{\pref}[1]{(\ref{#1})}
\newcommand{\figref}[1]{Fig.~\ref{#1}}

\begin{document}

\title{Direct Evidence for Universal Statistics of Stationary Kardar-Parisi-Zhang Interfaces}

\author{Takayasu Iwatsuka}
\affiliation{Department of Physics,\! Tokyo Institute of Technology,\! 2-12-1 Ookayama,\! Meguro-ku,\! Tokyo 152-8551,\! Japan}%
\affiliation{Department of Physics,\! The University of Tokyo,\! 7-3-1 Hongo,\! Bunkyo-ku,\! Tokyo 113-0033,\! Japan}%

\author{Yohsuke T. Fukai}
\affiliation{Nonequilibrium Physics of Living Matter RIKEN Hakubi Research Team,\! RIKEN Center for Biosystems Dynamics Research,\! 2-2-3 Minatojima-minamimachi,\! Chuo-ku,\! Kobe,\! Hyogo 650-0047,\! Japan}%
\affiliation{Department of Physics,\! The University of Tokyo,\! 7-3-1 Hongo,\! Bunkyo-ku,\! Tokyo 113-0033,\! Japan}%

\author{Kazumasa A. Takeuchi}
\email{kat@kaztake.org}
\affiliation{Department of Physics,\! The University of Tokyo,\! 7-3-1 Hongo,\! Bunkyo-ku,\! Tokyo 113-0033,\! Japan}%
\affiliation{Department of Physics,\! Tokyo Institute of Technology,\! 2-12-1 Ookayama,\! Meguro-ku,\! Tokyo 152-8551,\! Japan}%

\date{\today}

\begin{abstract}
The nonequilibrium steady state of the one-dimensional (1D) Kardar-Parisi-Zhang (KPZ) universality class is studied in-depth by exact solutions, yet no direct experimental evidence of its characteristic statistical properties has been reported so far. This is arguably because, for an infinitely large system, infinitely long time is needed to reach such a stationary state and also to converge to the predicted universal behavior. Here we circumvent this problem in the experimental system of growing liquid-crystal turbulence, by generating an initial condition that possesses a long-range property expected for the KPZ stationary state. The resulting interface fluctuations clearly show characteristic properties of the 1D stationary KPZ interfaces, including the convergence to the Baik-Rains distribution. We also identify finite-time corrections to the KPZ scaling laws, which turn out to play a major role in the direct test of the stationary KPZ interfaces. This paves the way to explore unsolved properties of the stationary KPZ interfaces experimentally, making possible connections to nonlinear fluctuating hydrodynamics and quantum spin chains as recent studies unveiled relation to the stationary KPZ.
\end{abstract}


\maketitle

\paragraph{Introduction.} 
The Kardar-Parisi-Zhang (KPZ) universality class
 describes dynamic scaling laws of a variety of phenomena,
 ranging from growing interfaces
 to directed polymers and stirred fluids
 \cite{Kardar.etal-PRL1986,Barabasi.Stanley-Book1995},
 as well as fluctuating hydrodynamics \cite{Spohn-Inbook2016} and, most recently, quantum integrable spin chains \cite{Ljubotina.etal-PRL2019,*Gopalakrishnan.Vasseur-PRL2019,*Das.etal-PRE2019}, to name but a few.
The KPZ class is now central in the studies of nonequilibrium scaling laws,
 mostly because some models in the one-dimensional (1D) KPZ class
 turned out to be integrable and exactly solvable
 (for reviews, see, e.g., \cite{Takeuchi-PA2018,Spohn-Inbook2017,*HalpinHealy.Takeuchi-JSP2015,*Corwin-RMTA2012}).
This has unveiled a wealth of nontrivial fluctuation properties
 in such nonequilibrium and nonlinear many-body problems.

The KPZ class is often characterized by the KPZ equation,
 a paradigmatic model for interfaces growing in fluctuating environments
 \cite{Kardar.etal-PRL1986,Barabasi.Stanley-Book1995,Takeuchi-PA2018}.
It reads, in the case of 1D interfaces in a plane:
\begin{equation}
 \prt{}{t}h(x,t) = \nu\prts{h}{x}{2} + \frac{\lambda}{2}\(\prt{h}{x}\)^2 + \eta(x,t).  \label{eq:KPZeq}
\end{equation}
Here $h(x,t)$ denotes the position of the interface
 in the direction normal to a reference line (e.g., substrate),
 often called the local height, at lateral position $x$ and time $t$.
$\eta(x,t)$ is white Gaussian noise with $\expct{\eta(x,t)}=0$ and
 $\expct{\eta(x,t)\eta(x',t')}=D\delta(x-x')\delta(t-t')$,
 where $\expct{\cdots}$ denotes the ensemble average.
Such random growth develops nontrivial fluctuations of $h(x,t)$,
 characterized by a set of universal power laws.
For example, the fluctuation amplitude of $h(x,t)$ grows as $t^\beta$,
 with $\beta=1/3$ for 1D.
This implies
\begin{equation}
 h(x,t) \simeq v_\infty t + (\Gamma t)^{1/3} \chi + \mathcal{O}(t^0)  \label{eq:Height}
\end{equation}
 with constant parameters $v_\infty,\Gamma$
 and a rescaled random variable $\chi$.
$\chi$ is correlated in space and time but characterized
 by a distribution that remains well defined in the limit $t\to\infty$.
Another important quantity is the height-difference correlation function,
 defined by $C_h(\ell,t)\equiv\expct{[h(x+\ell,t)-h(x,t)]^2}$.
While $C_h(\ell,t)\sim{}t^{2\beta}$ for $\ell$ much larger than
the correlation length $\xi(t)\sim{}t^{1/z}$,
 for $\ell\ll\xi(t)$, $C_h(\ell,t)\sim{}\ell^{\,2\alpha}$
 with $\alpha=z\beta$
 \cite{Barabasi.Stanley-Book1995,Takeuchi-PA2018}.
For 1D, the scaling exponents are $\alpha=1/2,\beta=1/3,z=3/2$
 and shared among members of the KPZ universality class
 \cite{Kardar.etal-PRL1986,Barabasi.Stanley-Book1995,Takeuchi-PA2018,Corwin-RMTA2012}.
Moreover, for the 1D KPZ equation \pref{eq:KPZeq}, 
 the (statistically) stationary state of this particular model,
 $h_\text{stat}^\text{KPZeq}(x)$,
 is known to be equivalent to the 1D Brownian motion
 \cite{Kardar.etal-PRL1986,Barabasi.Stanley-Book1995,Takeuchi-PA2018,Corwin-RMTA2012}:
\begin{equation}
 h_\text{stat}^\text{KPZeq}(x) = \sqrt{A}B(x).  \label{eq:1DBM}
\end{equation}
Here, $A\equiv{}D/2\nu$ and $B(x)$ is the standard Brownian motion
 with \textit{time} $x$,
 so that $\expct{B(x)}=0$ and $\expct{[B(x+\ell)-B(x)]^2}=\ell$.
The height-difference correlation function
 for $h_\text{stat}^\text{KPZeq}(x)$ is then simply
 the mean-squared displacement, $C_{h_\text{stat}^\text{KPZeq}}(\ell)\simeq{}A\ell$,
 with $A$ corresponding to the diffusion coefficient.
Note that, even if we set $h(x,0)=h_\text{stat}^\text{KPZeq}(x)$, $h(x,t)$ still fluctuates and grows, i.e., $\expct{h(x,t)} = v_\infty t$ with a constant $v_\infty$. 
Nevertheless, the shifted height $h(x,t) - v_\infty t$ can be always described by \eqref{eq:1DBM} with another instance of $B(x)$ (which is actually correlated with the one used for the initial condition).
For lack of a better term, here we call it the (statistically) stationary state of the KPZ equation.

Then the exact solutions of the 1D KPZ equation \cite{Sasamoto.Spohn-PRL2010,Amir.etal-CPAM2011,Calabrese.etal-EL2010,Dotsenko-EL2010,Calabrese.LeDoussal-PRL2011,Imamura.Sasamoto-PRL2012,Borodin.etal-MPAG2015}, as well as earlier results for discrete models (e.g., \cite{Johansson-CMP2000,Prahofer.Spohn-PRL2000}), unveiled detailed fluctuation properties of $h(x,t)$,
 in particular the distribution function of $\chi$
 \cite{Takeuchi-PA2018,Corwin-RMTA2012}.
Further, those properties turned out to depend
 on the global geometry of interfaces
 or on the initial condition $h(x,0)$,
 being classified into a few universality \textit{subclasses}
 within the single KPZ class.
Among them, most important and established are
 the subclasses for circular, flat, and stationary interfaces,
 characterized by the following asymptotic distributions
 \cite{Takeuchi-PA2018}:
 the GUE Tracy-Widom \cite{Tracy.Widom-CMP1994},
 GOE Tracy-Widom \cite{Tracy.Widom-CMP1996},
 and Baik-Rains distributions \cite{Baik.Rains-JSP2000},
 respectively (GUE and GOE stand for the Gaussian unitary and orthogonal ensembles, respectively).
More precisely, with the random numbers drawn from those distributions,
 denoted by $\chi_2,\chi_1,\chi_0$
\footnote{
With the standard GOE Tracy-Widom random variable $\chi_\text{1,TW}$
 (as defined in Ref.~\cite{Tracy.Widom-CMP1996}), $\chi_1$ is defined by
 $\chi_1\equiv2^{-2/3}\chi_\text{1,TW}$ \cite{Takeuchi-PA2018}.
}, respectively, we have
 $\chi\tod\chi_2,\chi_1,\chi_0$ for the three respective subclasses
 \footnote{
In the circular case, for $x \neq 0$, an additional shift proportional to $x^2/t$ is needed for the convergence to $\chi_2$, to compensate the locally parabolic mean profile of the interfaces \cite{Takeuchi-PA2018,Spohn-Inbook2017}.
In the stationary case, the left-hand side of \eqref{eq:Height} should be more precisely $h(x,t)-h(x,0)$, but by imposing $h(0,0)=0$ one can still use \eqref{eq:Height} at $x=0$ to show $\chi \tod \chi_0$ \cite{Takeuchi-PA2018,Spohn-Inbook2017}.
},
 where $\tod$ indicates the convergence in the distribution.
For the KPZ equation, the typical initial conditions
 that correspond to the three subclasses are
 $h(x,0)=-|x|/\delta~(\delta\to0^+)$ (circular),
 $h(x,0)=0$ (flat),
 and $h(x,0)=h_\text{stat}^\text{KPZeq}(x)=\sqrt{A}B(x)$ (stationary).
Experimentally, the circular and flat subclasses were clearly observed
 in the growth of liquid-crystal turbulence
 \cite{Takeuchi.Sano-PRL2010,*Takeuchi.etal-SR2011,Takeuchi.Sano-JSP2012,Takeuchi-PA2018},
 but only indirect and partial support has been reported so far
 for the stationary subclass \cite{Takeuchi-PRL2013,Takeuchi-JPA2017}
 (see also 
\footnote{
There was a claim for an observation of the Baik-Rains distribution
 in an experiment of paper combustion \cite{Miettinen.etal-EPJB2005},
 but it seems to us that their precision is not sufficient to distinguish it
 from other possible distributions,
 as detailed in the commentary article available
 at \protect\url{http://publ.kaztake.org/miet-com.pdf}. 
Note also that in the stationary state of a finite-size system,
 as studied in this experiment, an approach to the Baik-Rains distribution
 will appear in a finite time window,
 so that careful analysis of time dependence is crucial.
}).
This is presumably because, firstly,
 for an infinitely large system,
 it takes infinitely long time for a system to reach the stationary state
 (as $\xi(t)\sim{}t^{2/3}$ needs to reach infinity).
Then one should take an interface profile in the stationary state,
 regard it as an ``initial condition'', and wait sufficiently long time
 for the height fluctuations to converge to the Baik-Rains distribution
 (see Ref.~\cite{Takeuchi-PRL2013} for more quantitative arguments).
For a finite system of size $L$,
 reaching the stationary state takes a finite time $\sim{}L^{3/2}$,
 but the approach to the Baik-Rains distribution is now visible
 only within a finite time period \cite{Prolhac-PRL2016,Baik.Liu-a2016}, being eventually replaced by a final state unrelated to the choice of the initial condition.

\begin{figure}[btp]
 \centering
 \includegraphics[clip]{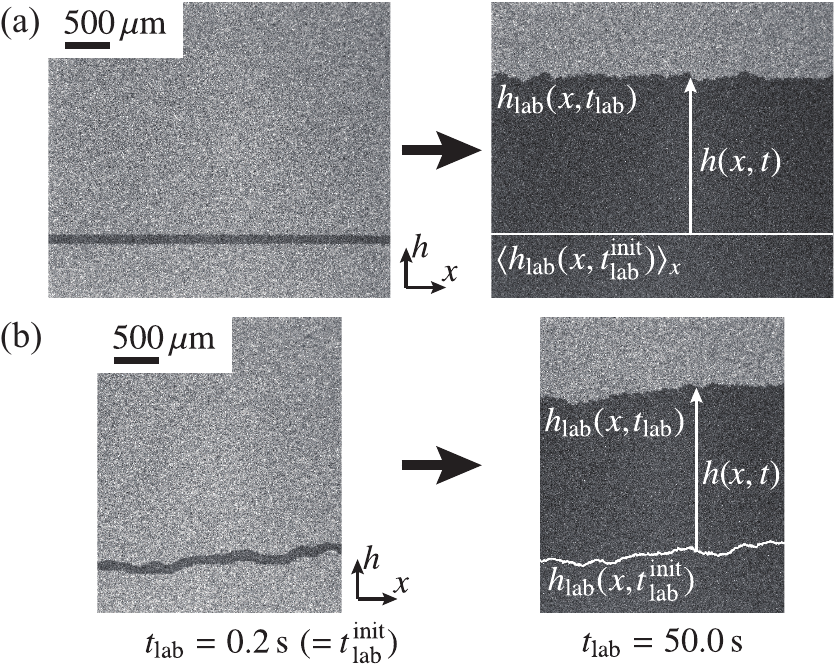}
 \caption{
Typical snapshots of a flat~(a) and a Brownian~(b) interface,
 separating the metastable DSM1 (gray) and growing DSM2 regions (black).
$h_\text{lab}(x,t_\text{lab})$ denotes the position of the upper interface in the laboratory frame, at time $t_\text{lab}$ from the laser emission.
$t$ and $h(x,t)$ are defined as follows: $t\equiv{}t_\text{lab}$ and $h(x,t)\equiv{}h_\text{lab}(x,t_\text{lab})-\expct{h(x,t_\text{lab}^\text{init})}_x$ for the flat case (a), $t\equiv{}t_\text{lab}-t_\text{lab}^\text{init}$ and $h(x,t)\equiv{}h(x,t_\text{lab})-h(x,t_\text{lab}^\text{init})$ for the Brownian case (b).
See also Movies~S1 and S2 \cite{SM}.
}
 \label{fig1}
\end{figure}

Here we overcome this difficulty in the liquid-crystal experimental system,
 by \textit{generating} an interface
 that resembles the expected stationary state.
Using a holographic technique developed previously
 \cite{Fukai.Takeuchi-PRL2017,*Fukai.Takeuchi-PRL2020},
 we generated Brownian initial conditions \pref{eq:1DBM}
 for the growing turbulence
 and directly measured fluctuation properties of the height $h(x,t)$
 under this type of initial conditions [\figref{fig1}(b)].
This allowed us to carry out quantitative tests of a wealth of exact results
 for integrable models in the stationary state.
And indeed, we obtained direct evidence for the Baik-Rains distribution and the related correlation function.
This opens an experimental pathway to explore universal yet hitherto unsolved statistical properties of the KPZ stationary state.

\paragraph{Methods.}
The experimental system was a minor modification of that
 used in Ref.~\cite{Fukai.Takeuchi-PRL2017,*Fukai.Takeuchi-PRL2020}
 (see Sec.~I of Supplementary Text and Fig.~S1 \cite{SM} for details).
We used a standard material for the electroconvection
 of nematic liquid crystal \cite{deGennes.Prost-Book1995}, specifically,
 $N$-(4-methoxybenzylidene)-4-butylaniline
 doped with tetra-$n$-butylammonium bromide.
The liquid crystal sample was placed between two parallel glass plates
 with transparent electrodes,
 separated by spacers of thickness $12\unit{\mu{}m}$.
The electrodes were surface-treated to realize homeotropic alignment.
The temperature was maintained at $25\unit{^\circ{}C}$
 during the experiments, with typical fluctuations of $0.01\unit{^\circ{}C}$.

The electroconvection was induced by applying an ac voltage to the system.
In this work we fixed the frequency at $250\unit{Hz}$,
 well below the cut-off frequency near $1.8\unit{kHz}$,
 and the voltage was set to be $23\unit{V}$.
At this voltage, the system is initially in a turbulent state
 called the dynamic scattering mode 1 (DSM1), which is actually metastable,
 so that the stable turbulent state DSM2 eventually nucleates and expands,
 forming a growing cluster bordered by a fluctuating interface.
One can also trigger DSM2 nucleation
 by shooting an ultraviolet (UV) laser pulse \cite{Takeuchi-PA2018}.
This not only allows us to carry out controlled experiments
 but also to design the initial shape of the interface,
 by changing the intensity profile of the laser beam.
Growing interfaces were observed by recording light transmitted
 through the sample, using a light-emitting diode as the light source
 and a charge-coupled device camera.

\paragraph{Flat interface experiments.}

In order to realize Brownian initial conditions \pref{eq:1DBM}
 that may correspond to the stationary state,
 we first need to evaluate the parameter $A$.
To this end we first carried out a set of experiments for flat interfaces.
Using a cylindrical lens to expand the laser beam,
 we generated an initially straight interface for each experiment
 and tracked growth of the upper interface [\figref{fig1}(a)].
The $h$-axis is set along the mean growth direction.
The $x$-axis is normal to $h$, along the initial straight line.
Then the coordinates of the upper interface in the laboratory frame were extracted and denoted by $h_\text{lab}(x,t_\text{lab})$, where $t_\text{lab}$ is the time elapsed since the laser emission.
Since the height of interest is the increment from the initial interface, we approximated it by the spatially averaged height at the first analyzable time, denoted by $\expct{h(x,t_\text{lab}^\text{init})}_x$, with $t_\text{lab}^\text{init}=0.2\unit{s}$.
Then we defined $h(x,t)\equiv{}h_\text{lab}(x,t_\text{lab})-\expct{h(x,t_\text{lab}^\text{init})}_x$ with $t\equiv{}t_\text{lab}$ and studied its fluctuations over 1267 independent realizations.
In the following, the ensemble average $\expct{\cdots}$ was evaluated by averaging over all realizations and spatial points $x$.

\begin{figure}[btp]
 \centering
 \includegraphics[clip,width=\hsize]{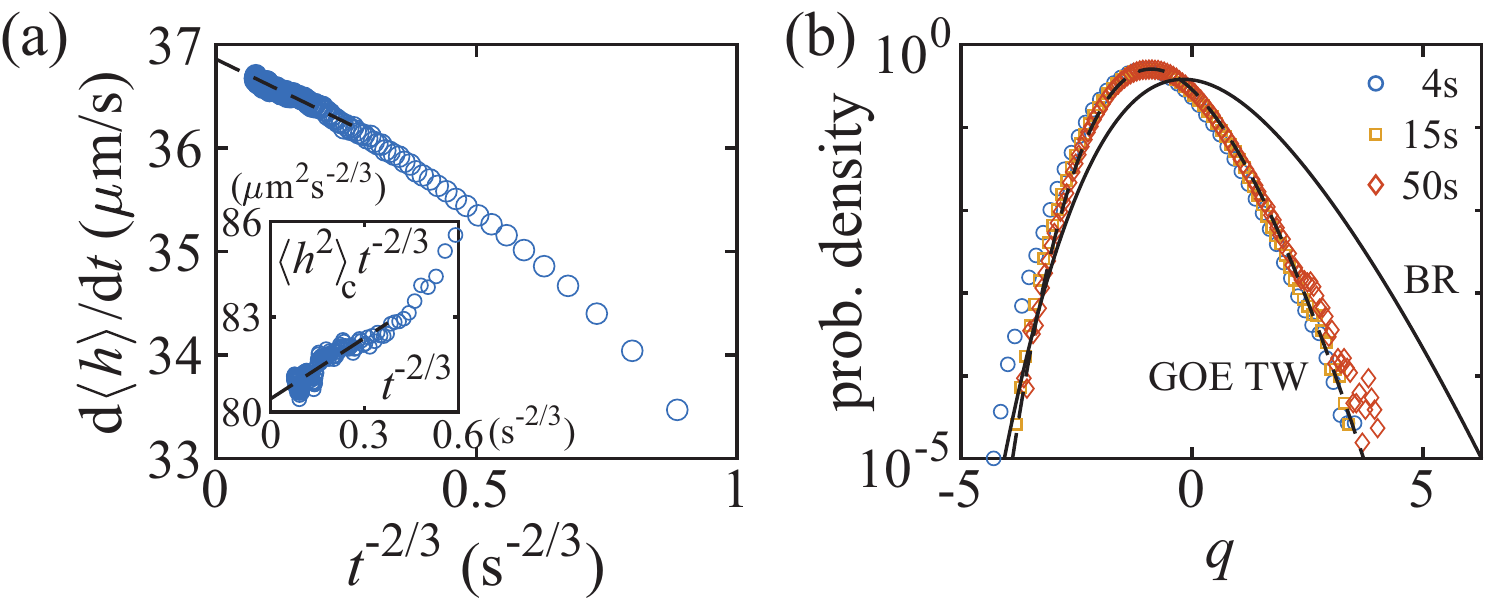}
 \caption{
Parameter estimation for the flat interfaces.
(a) $\diff{\expct{h}}{t}$ against $t^{-2/3}$ (main panel) and $\cum{h^2}t^{-2/3}$ against $t^{-2/3}$ (inset).
The dashed lines show the results of linear regression.
(b) Histograms of the rescaled height $q(x,t)$ at different $t$ (legend).
Agreement with the GOE Tracy-Widom (TW) distribution is confirmed.
BR stands for the Baik-Rains distribution.
}
 \label{fig2}
\end{figure}

The parameter $A$ can be determined by the relation $A=\sqrt{2\Gamma/v_\infty}$, known to hold in isotropic systems \cite{Takeuchi.Sano-JSP2012,Takeuchi-PA2018}.
For $v_\infty$, we followed the standard procedure \cite{Krug.etal-PRA1992,Takeuchi-PA2018} and plotted $\diff{\expct{h}}{t}$ against $t^{-2/3}$ [\figref{fig2}(a) main panel].
From \eqref{eq:Height}, we have
\begin{equation}
 \diff{\expct{h}}{t} \simeq v_\infty + \frac{\Gamma^{1/3}\expct{\chi}}{3} t^{-2/3}.  \label{eq:dhdt}
\end{equation}
Therefore, reading the $y$-intercept of linear regression, we obtained $v_\infty=36.86(4)\unit{\mu{}m/s}$, where the numbers in the parentheses indicate the uncertainty.
For $\Gamma$, since the flat interfaces in this liquid-crystal system were already shown to exhibit the GOE Tracy-Widom distribution \cite{Takeuchi.etal-SR2011,Takeuchi.Sano-JSP2012,Takeuchi-PA2018}, we have $\cum{h^n}\simeq(\Gamma{}t)^{n/3}\cum{\chi_1^n} (n\geq2)$, where $\cum{X^n}$ denotes the $n$th-order cumulant of a variable $X$.
Above all, the variance can be most precisely determined, and is known to grow, with the leading finite-time correction, as $\cum{h^2}\simeq(\Gamma{}t)^{2/3}\cum{\chi_1^2}+\mathcal{O}(t^0)$ \cite{Takeuchi.Sano-JSP2012,Takeuchi-PA2018,Ferrari.Frings-JSP2011}.
Therefore, by plotting $\cum{h^2}t^{-2/3}$ against $t^{-2/3}$ [\figref{fig2}(a) inset] and reading the $y$-intercept of linear regression, we obtained $\Gamma=1415(4)\unit{\mu{}m^3/s}$
\footnote{
Although $\Gamma$ can also be estimated from \eqref{eq:dhdt}, reading the slope of \figref{fig2}(a) is much less precise than the estimation based on the variance.
}.
Consistency was checked by plotting the histogram of the height, rescaled with those parameters as follows
\begin{equation}
    q(x,t) \equiv \frac{h(x,t) - v_\infty t}{(\Gamma t)^{1/3}} \simeq \chi.  \label{eq:q}
\end{equation}
Clear agreement with the GOE Tracy-Widom distribution was confirmed [\figref{fig2}(b)].
Using those estimates, we finally obtained $A=\sqrt{2\Gamma/v_\infty}=8.762(13)\unit{\mu{}m}$.

\paragraph{Brownian interface experiments.}

Based on the value of $A$ evaluated by the flat interface experiments, we generated Brownian initial conditions \pref{eq:1DBM} with $A=9\unit{\mu{}m}$ 
\footnote{
To reduce the effect of parameter shift, we chose to start the Brownian interface experiments before completing the careful analysis of the flat experimental data.
As a result, we used a rough estimate $A=9\unit{\mu{}m}$ for the Brownian initial conditions.
A slight difference in $A$ is expected to have only a minor impact on the fluctuation properties of $h(x,t)$ \cite{Chhita.etal-AAP2018}.
}
and studied growing DSM2 interfaces [\figref{fig1}(b)].
Each initial condition was prepared by projecting a hologram of a computer-generated Brownian trajectory, with resolution of $36.5\unit{\mu{}m}$ at the liquid-crystal cell, by using a spatial light modulator \cite{SM}.
The height profile in the laboratory frame $h_\text{lab}(x,t_\text{lab})$ was determined as for the flat experiments, but here the height of interest is the increment from the height profile at the first analyzable time, $h(x,t)\equiv{}h_\text{lab}(x,t_\text{lab})-h_\text{lab}(x,t_\text{lab}^\text{init})$, with $t\equiv{}t_\text{lab}-t_\text{lab}^\text{init}$ and $t_\text{lab}^\text{init}=0.2\unit{s}$ [\figref{fig1}(b)].
We used a region of width $2730\unit{\mu{}m}$ near the center of the camera view and analyzed 1021 interfaces.
Finite-size effect is expected to be prevented, because the Brownian trajectories were much longer ($4670\unit{\mu{}m}$ in $x$) than the width of the analyzed region.

\begin{figure}[btp]
 \centering
 \includegraphics[clip,width=\hsize]{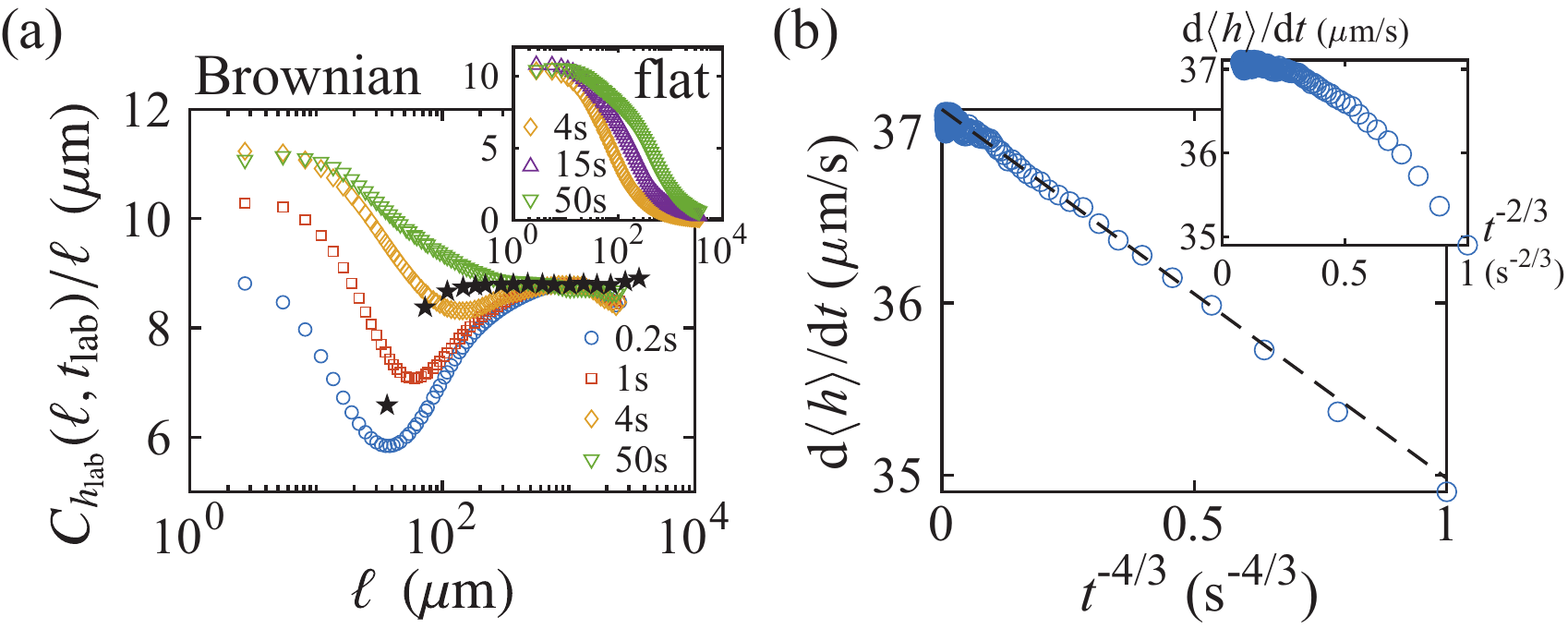}
 \caption{
Evaluation of the Brownian interfaces.
(a) $C_{h_\text{lab}}(\ell,t_\text{lab})/\ell$ against $\ell$ for different $t_\text{lab}$ (indicated in the legends) for the Brownian (main panel) and flat (inset) interfaces.
The black stars indicate the results of direct evaluation of the computer-generated images used for the holograms.
(b) $\diff{\expct{h}}{t}$ against $t^{-4/3}$ (main panel) and $t^{-2/3}$ (inset) for the Brownian interfaces.
The dashed line in the main panel shows the result of linear regression.
}
 \label{fig3}
\end{figure}

First we test whether the interfaces generated thereby are stationary or not.
To this end, we measure the height-difference correlation function for $h_\text{lab}(x,t_\text{lab})$, $C_{h_\text{lab}}(\ell,t_\text{lab})$, and find that it does depend on $t_\text{lab}$ [\figref{fig3}(a)], indicating that the interfaces are \textit{not} stationary.
More precisely, we observe that $C_{h_\text{lab}}(\ell,t_\text{lab})/\ell$ at small $\ell$ initially takes values lower than the desired one, $A=9\unit{\mu{}m}$, presumably because of the finite resolution of the holograms, then increases up to $\approx{}11\unit{\mu{}m}$.
The fact that $C_{h_\text{lab}}(\ell,t_\text{lab})/\ell$ becomes higher than $A$ at small $\ell$ was also observed in our flat data [\figref{fig3}(a) inset] as well as in our past experiments \cite{Takeuchi.Sano-PRL2010,Takeuchi-PA2018}.
However, more important is the behavior at large $\ell$, which turns out to be stable and takes a value close to $A=9\unit{\mu{}m}$.
Therefore, in the following we test whether our interfaces, though not stationary, can nevertheless exhibit universal properties of the stationary KPZ subclass, such as the Baik-Rains distribution.

To determine the scaling coefficients, we plot 
$\diff{\expct{h}}{t}$ against $t^{-2/3}$ in the inset of \figref{fig3}(b).
Time dependence of $\diff{\expct{h}}{t}$ confirms non-stationarity of the interfaces again.
Interestingly, as opposed to the result for the flat interfaces [\figref{fig2}(a)], here we do not find linear relationship to $t^{-2/3}$ [\figref{fig3}(b) inset], but to $t^{-4/3}$ (main panel).
From \eqref{eq:dhdt}, this suggests $\expct{\chi}=0$, consistent with the vanishing mean of the Baik-Rains distribution $\expct{\chi_0}=0$.
If so, the subleading term of \eqref{eq:dhdt} is indeed expected to be $\mathcal{O}(t^{-4/3})$, coming from a $t^{-1/3}$ term expected to exist in \eqref{eq:Height}.
Then, by linear regression, we obtained $v_\infty=37.126(15)\unit{\mu{}m/s}$. 
It is reasonably close to the value from the flat experiments, in view of the typical magnitude of parameter shifts in this experimental system \cite{Takeuchi.Sano-JSP2012}.
For $\Gamma$, we took the value from the flat experiments, so that we do not make any assumption on the statistical properties for the Brownian case.

\begin{figure}[btp]
 \centering
 \includegraphics[clip,width=\hsize]{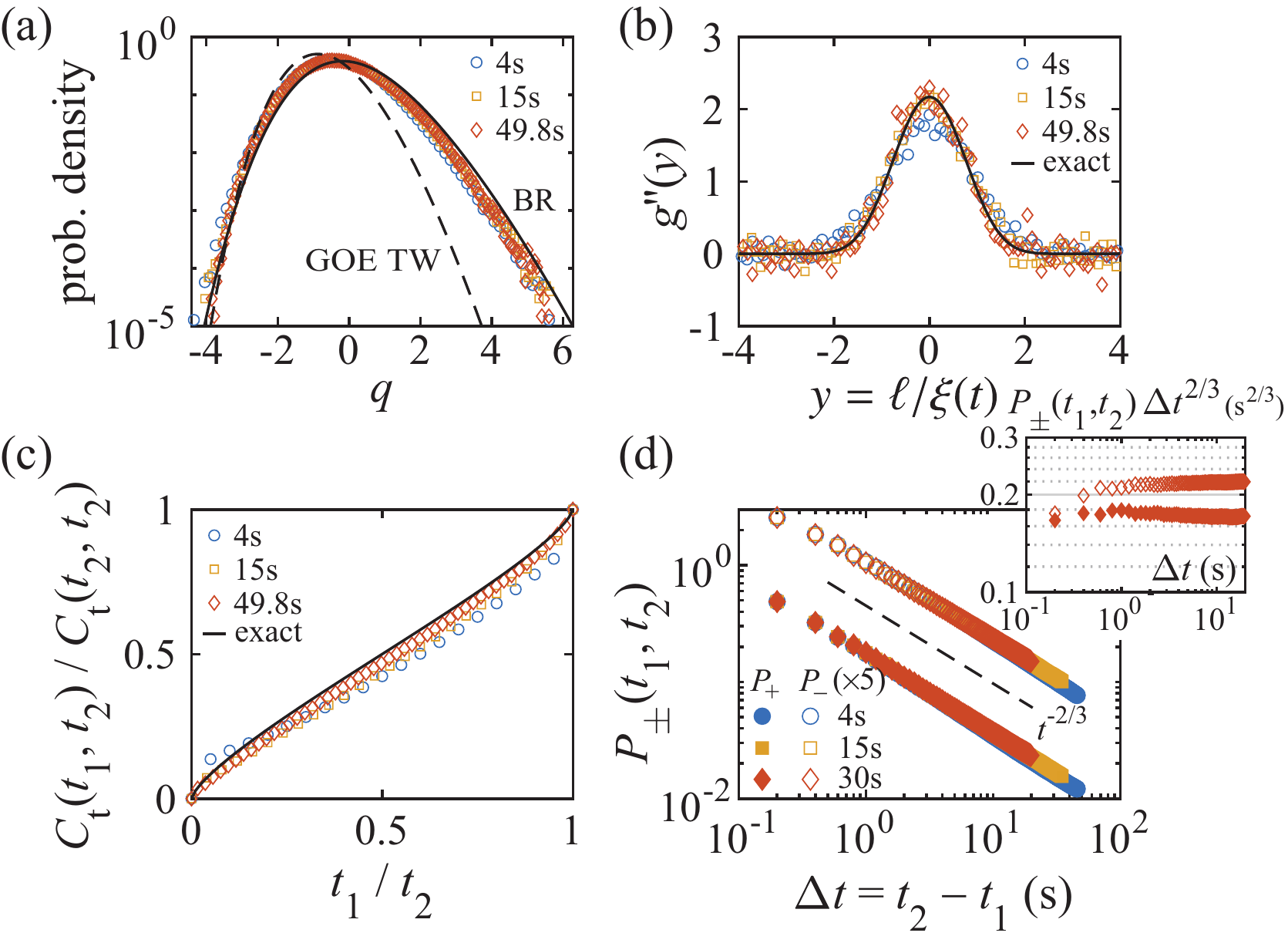}
 \caption{
Main results of the Brownian interface experiments.
(a) Histograms of the rescaled height $q(x,t)$ at different $t$ (legend).
The data are found to converge to the Baik-Rains (BR) distribution, as shown quantitatively in Sec.~II of Supplementary Text and Fig.~S2 \cite{SM}.
GOE TW stands for the GOE Tracy-Widom distribution.
(b) Two-point correlation function $g''(y)$.
The experimental data are evaluated by $\frac{\xi(t)^2}{(\Gamma{}t)^{2/3}}2\expct{\prt{\hlab}{x}(\ell+x,t+\tlabinit)\prt{\hlab}{x}(x,\tlabinit)}$ with different $t$ (legend).
The black curve indicates Pr\"ahofer and Spohn's exact solution \cite{Prahofer.Spohn-JSP2004,Prahofer.Spohn-Table}.
(c) Rescaled two-time function $C_\mathrm{t}(t_1,t_2)/C_\mathrm{t}(t_2,t_2)$ for different $t_2$ (legend).
The data are found to converge to Ferrari and Spohn's exact solution \cite{Ferrari.Spohn-SIG2016} (black curve), as shown quantitatively in Sec.~III of Supplementary Text and Fig.~S3 \cite{SM}.
(d) Persistence probability $P_\pm(t_1,t_2)$ for different $t_1$ (legend).
For visibility, $P_-(t_1,t_2)$ is shifted by factor $5$.
The dashed line is a guide for eyes indicating the power law $t^{-2/3}$ for \FBM.
The inset shows $P_\pm(t_1,t_2)\Delta{}t^{2/3}$.
}
 \label{fig4}
\end{figure}%

Using the values of $v_\infty$ and $\Gamma$ determined thereby, as well as $A=\sqrt{2\Gamma/v_\infty}$, we test various predictions for the stationary KPZ subclass, without any adjustable parameter.
The results are summarized in \figref{fig4}.
Figure~\ref{fig4}(a) shows histograms of the rescaled height $q(x,t)$ [\eqref{eq:q}] at different times $t$.
The obtained distributions at finite times are already close to the predicted Baik-Rains distribution.
Indeed, convergence in the $t\to\infty$ limit is confirmed quantitatively by analyzing finite-time corrections in the cumulants (Sec.~II of Supplementary Text and Fig.~S2 \cite{SM}).
In \figref{fig4}(b), we test the prediction on the two-point correlation function $C_2(\ell,t)\equiv\expct{[\hlab(\ell+x,t+t_0)-\hlab(x,t_0)-v_\infty{}t]^2}$.
It is often denoted by $g(y)$ in the rescaled units, with $y\equiv\ell/\xi(t)$, $\xi(t)\equiv(2/A)(\Gamma{}t)^{2/3}$, and $g(y)\equiv(\Gamma{}t)^{-2/3}C_2(\ell,t)$.
Its second derivative, $g''(y)$, plays the pivotal role in the emergence of KPZ in fluctuating hydrodynamics \cite{Spohn-Inbook2016} and quantum integrable spin chains \cite{Ljubotina.etal-PRL2019}.
This is tested with our experimental data and good agreement is found [\figref{fig4}(b)].
Figure~\ref{fig4}(c) shows the results of the two-time correlation of $h(x,t)$, $C_\mathrm{t}(t_1,t_2)\equiv\expct{\delta{}h(x,t_1)\delta{}h(x,t_2)}$ with $\delta{}h(x,t)\equiv{}h(x,t)-\expct{h(x,t)}$.
Our data agree with Ferrari and Spohn's prediction \cite{Ferrari.Spohn-SIG2016} that the two-time correlation coincides with that of the fractional Brownian motion with Hurst exponent $1/3$ (hereafter abbreviated to \FBM), $C_\mathrm{t}(t_1,t_2)/C_\mathrm{t}(t_2,t_2)\to(1/2)[1+(t_1/t_2)^{2/3}-(1-t_1/t_2)^{2/3}]$ (black line) in the limit $t_1,t_2\to\infty$ with fixed $t_1/t_2$ (see Sec.~III of Supplementary Text and Fig.~S3 \cite{SM} for a quantitative test).
Finally, \figref{fig4}(d) shows the persistence probability $P_\pm(t_1,t_2)$, i.e., the probability that $h(x,t)-h(x,t_1)$ remains always positive ($P_+$) or negative ($P_-$) until time $t_2$, which is found to decay clearly as $P_\pm(t_1,t_2)\sim{}\Delta{}t^{-2/3}$ with $\Delta{}t\equiv{}t_2-t_1$.
The persistence exponent is therefore $2/3$, supporting Krug \textit{et al.}'s conjecture \cite{Krug.etal-PRE1997,*Kallabis.Krug-EL1999,*Bray.etal-AP2013} that it also coincides with that of \FBM. 
Those relations to \FBM\, are intriguing, because $h(x,t)$ is not Gaussian and therefore its time evolution is \textit{not} \FBM.

\paragraph{Concluding remarks.}

In this work we aimed at unambiguous tests of universal statistics for the stationary state of the $(1+1)$-dimensional KPZ class.
Instead of waiting for the interfaces to approach the stationary state, we generated such initial conditions that are expected to share the same long-range properties with the stationary state, specifically, the Brownian initial conditions \pref{eq:1DBM} with the appropriate diffusion coefficient $A$ determined beforehand.
The resulting interfaces turned out to be \textit{not} stationary, but nevertheless our data clearly showed the defining properties of the stationary KPZ subclass, including the Baik-Rains distribution and the two-point correlation function $g''(y)$ [\figref{fig4}(a)(b)].
Our results also support intriguing relations to time correlation properties of the fractional Brownian motion [\figref{fig4}(c)(d)], which may deserve further investigations in other quantities.
With this and past studies \cite{Takeuchi.Sano-PRL2010,Takeuchi.etal-SR2011,Takeuchi.Sano-JSP2012,Takeuchi-PA2018},  all the three representative KPZ subclasses in one dimension \cite{Takeuchi-PA2018,Corwin-RMTA2012} were given experimental supports for the universality.

The KPZ class has been extensively studied already for decades, yet it continues finding novel connections to various areas of physics (recall recent developments in nonlinear fluctuating hydrodynamics \cite{Spohn-Inbook2016} and quantum spin chains \cite{Ljubotina.etal-PRL2019}).
We hope our experiments will also serve to probe quantities of interest for those systems, which may be not always solved exactly but still have a possibility to be measured precisely.
Explorations of higher dimensions, for which numerics have played leading roles \cite{HalpinHealy-PRL2012,*HalpinHealy-PRE2013,*Oliveira.etal-PRE2013,HalpinHealy.Takeuchi-JSP2015}, are also important directions left for future studies.

\begin{acknowledgments}
\paragraph{Acknowledgments.} 
We thank P. L. Ferrari, T. Halpin-Healy, T. Sasamoto, and H. Spohn for enlightening discussions. 
We are also grateful to M. Pr\"{a}hofer and H. Spohn
 for the theoretical curves of the BR and GOE-TW distributions
 and that of the stationary correlation function $g(\zeta)$,
 which are made available online \cite{Prahofer.Spohn-Table}.
This work is supported in part by KAKENHI
 from Japan Society for the Promotion of Science
 (Grant Nos. JP25103004, JP16H04033, JP19H05144, JP19H05800, JP20H01826, JP17J05559),
 by Tokyo Tech Challenging Research Award 2016, by Yamada Science Foundation, 
 and by the National Science Foundation (Grant No. NSF PHY11-25915). 
\end{acknowledgments}

\bibliography{IwaTak}

\end{document}


\title{Supplementary Information for \\ ``Direct Evidence for Universal Statistics of Stationary Kardar-Parisi-Zhang Interfaces''}

\author{Takayasu Iwatsuka}
\affiliation{Department of Physics,\! Tokyo Institute of Technology,\! 2-12-1 Ookayama,\! Meguro-ku,\! Tokyo 152-8551,\! Japan}%
\affiliation{Department of Physics,\! The University of Tokyo,\! 7-3-1 Hongo,\! Bunkyo-ku,\! Tokyo 113-0033,\! Japan}%

\author{Yohsuke T. Fukai}
\affiliation{Nonequilibrium Physics of Living Matter RIKEN Hakubi Research Team,\! RIKEN Center for Biosystems Dynamics Research,\! 2-2-3 Minatojima-minamimachi,\! Chuo-ku,\! Kobe,\! Hyogo 650-0047,\! Japan}%
\affiliation{Department of Physics,\! The University of Tokyo,\! 7-3-1 Hongo,\! Bunkyo-ku,\! Tokyo 113-0033,\! Japan}%

\author{Kazumasa A. Takeuchi}
\email{kat@kaztake.org}
\affiliation{Department of Physics,\! The University of Tokyo,\! 7-3-1 Hongo,\! Bunkyo-ku,\! Tokyo 113-0033,\! Japan}%
\affiliation{Department of Physics,\! Tokyo Institute of Technology,\! 2-12-1 Ookayama,\! Meguro-ku,\! Tokyo 152-8551,\! Japan}%

\date{\today}

\maketitle

\section{I. Experimental setup}

The experimental setup we used was a minor modification of the system used for our past studies \cite{Fukai.Takeuchi-PRL2017,*Fukai.Takeuchi-PRL2020} (see also \cite{Takeuchi.Sano-JSP2012}).
The convection cell consisted of a nematic liquid crystal sample, sandwiched between two parallel glass plates with transparent electrodes.
The material was $N$-(4-methoxybenzylidene)-4-butylaniline (TCI Chemicals), doped with 0.01~wt.\% of tetra-$n$-butylammonium bromide.
It was confined between two parallel glass plates coated with transparent electrodes (indium tin oxide), separated by spacers of thickness $12\unit{\mu{}m}$, which enclosed an observation area of roughly $17\unit{mm} \times 17\unit{mm}$.
The electrodes were coated with $N$,$N$-dimethyl-$N$-octadecyl-3-aminopropyltrimethoxysilyl chloride to realize the homeotropic alignment.
During the experiments, the temperature of the convection cell was kept constant at $25\unit{^\circ{}C}$, by using a hand-made thermocontroller and a thermally insulating chamber.
The thermally insulating chamber encloses the entire experimental setup and stabilizes the temperature inside roughly, by circulating water of a constant temperature.
The thermocontroller contains the convection cell and operates by a feedback-controlled Peltier element.
As a result, typical fluctuations of the temperature inside the thermocontroller were $0.01\unit{^\circ{}C}$ or smaller.

\begin{figure}[tb]
 \centering
 \includegraphics[clip,width=\hsize]{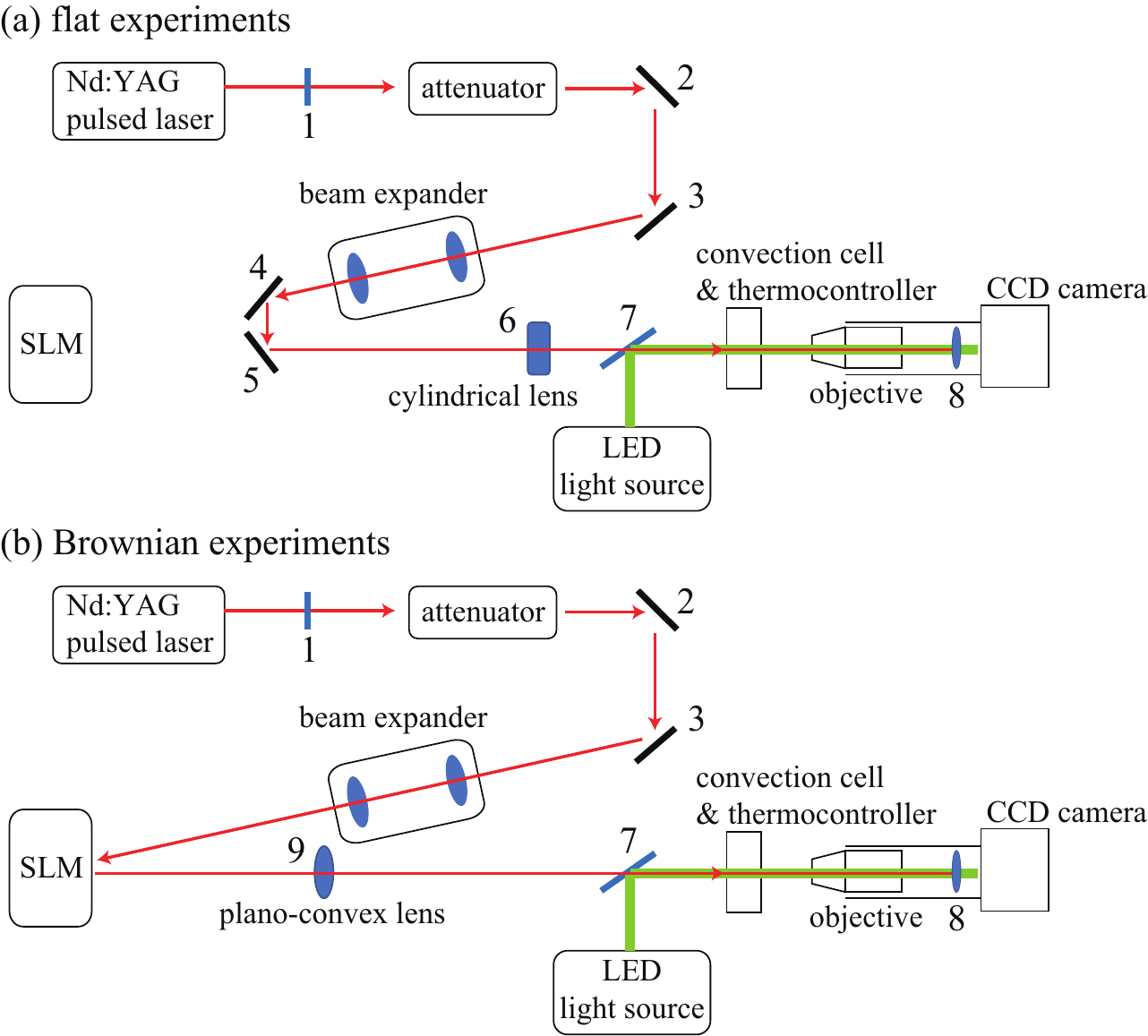}
 \caption{
Schematics of the optical systems for the flat (a) and Brownian (b) experiments.
1: ultraviolet band-pass filter.
2-5: mirror.
6: cylindrical lens.
7: dichroic mirror.
8: plano-convex lens and dichroic mirror (visible light pass).
9: plano-convex lens.
Nd:YAG pulsed laser (MiniLase, New Wave Research).
SLM: spatial light modulator (LCOS-SLM, X10468-05, Hamamatsu).
LED: light-emitting diode.
Convection cell: see the text.
Thermocontroller: hand-made, operating by feedback control of a Peltier element.
Objective (UplanFLN4x, Olympus).
CCD camera: charge-coupled device camera.
Note that the entire setup is contained in a thermally insulating chamber and the temperature inside is controlled to be constant.
}
 \label{fig-exp}
\end{figure}%

In the present work, we generated the growing DSM2 turbulence by shooting ultraviolet laser pulses to the convection cell, in the shape of a straight line (flat experiments) or a Brownian trajectory (Brownian experiments).
The schematics of the optical systems are shown in \figref{fig-exp}(a) and (b), respectively.
In both cases, the third harmonics ($355\unit{nm}$) of the Nd:YAG laser (MiniLase, New Wave Research) was used.
After the beam was attenuated and expanded, for the flat experiments, it was focused on a line by a cylindrical lens [6 in \figref{fig-exp}(a)].
For the Brownian experiments, the beam was sent to a spatial light modulator (LCOS-SLM, X10468-05, Hamamatsu) and to a plano-convex lens (9) to generate a holographic image on the focal plane [\figref{fig-exp}(b)].

\section{II. Finite-time corrections}

\begin{figure}[tb]
 \centering
 \includegraphics[clip,width=\hsize]{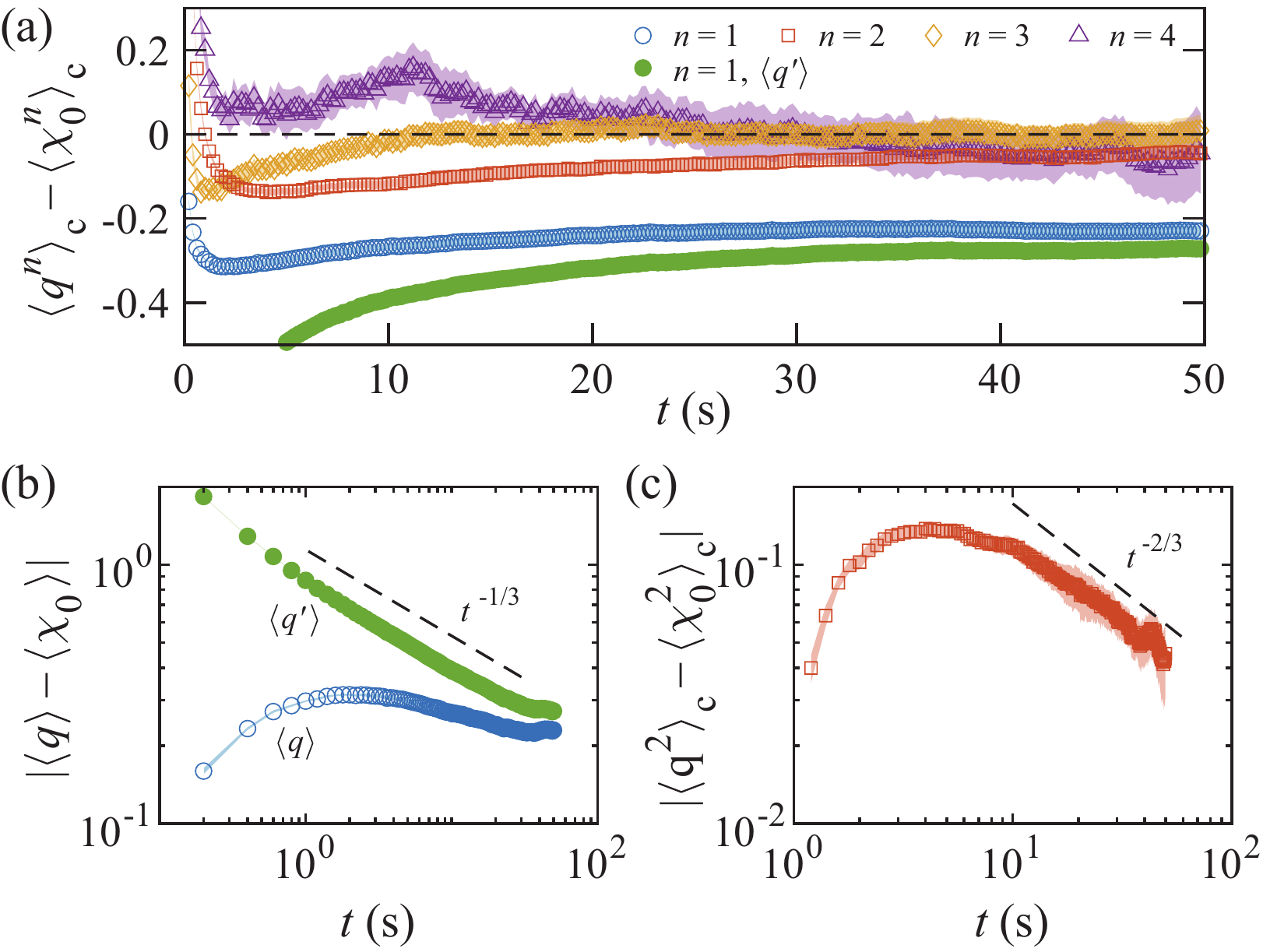}
 \caption{
Results on finite-time corrections to the Baik-Rains distribution for the Brownian interfaces.
(a) Difference between the $n$th-order cumulant of the rescaled height, $\cum{q^n}$, and that of the Baik-Rains distribution, $\cum{\chi_0^n}$, as functions of time. For the green solid disks, the mean of the refined rescaled height $q'$ [\eqref{eq:qprime}] is used instead.
The shades indicate the standard errors, evaluated by dividing the set of the realizations into 10 groups and computing the cumulants in each group.
The dashed line indicates zero, i.e., agreement with the Baik-Rains distribution.
}
 \label{fig-cum}
\end{figure}

Here we test agreement with the Baik-Rains distribution quantitatively, by analyzing time dependence of the cumulants of the rescaled height.
The rescaled height $q(x,t)$ is defined by 
\begin{equation}
    q(x,t) \equiv \frac{h(x,t) - v_\infty t}{(\Gamma t)^{1/3}} \simeq \chi + \mathcal{O}(t^{-1/3}).  \label{eq:q}
\end{equation}
Figure~\ref{fig-cum}(a) shows the difference between its $n$th-order cumulant $\cum{q^n}$ and that of the Baik-Rains distribution, $\cum{\chi_0^n}$, as functions of time, up to $n=4$.
We can see that the data for the third- and fourth-order cumulants (yellow diamonds and purple triangles, respectively) agree with those of the Baik-Rains distribution at late times, within the range of statistical errors (shades).
In contrast, the mean $\expct{q}$ (blue open circles) and the variance $\cum{q^2}$ (red squares) do not reach $\expct{\chi_0}$ and $\cum{\chi_0^2}$, respectively, though the difference is decreasing with increasing time.
Those differences are plotted in \figref{fig-cum}(b) and (c), respectively, in the log-log scales, with the same colors and symbols.
For the variance, we find $\cum{q^2} \simeq \cum{\chi_0^2} + \mathcal{O}(t^{-2/3})$ [\figref{fig-cum}(c)], which indicates convergence to the Baik-Rains variance in the limit $t\to\infty$.
Note that the finite-time correction of the variance was previously studied for the circular and flat KPZ subclasses, both experimentally \cite{Takeuchi.Sano-JSP2012} and theoretically \cite{Ferrari.Frings-JSP2011}, and the same exponent was obtained.

In contrast, the finite-time correction in the mean $\expct{q}$ [blue open circles in \figref{fig-cum}(b)] seems to show unusual behavior, decreasing significantly more slowly than the power law $t^{-1/3}$ expected from \eqref{eq:q}.
In fact, this can be understood by considering the next subleading term.
Suppose
\begin{equation}
 \expct{h(x,t)} \simeq v_\infty t + (\Gamma t)^{1/3} \expct{\chi} + A_1 + A_2 t^{-1/3}  \label{eq:Mean}
\end{equation}
The coefficient $A_2$ can be evaluated by the time dependence of the growth speed $\diff{\expct{h}}{t}$.
Hypothesizing that $\expct{\chi} = \expct{\chi_0} = 0$, we have
\begin{equation}
 \diff{\expct{h}}{t} \simeq v_\infty - \frac{A_2}{3}t^{-4/3}.  \label{eq:dhdt}
\end{equation}
This is exactly what we have seen in Fig.~3(b).
Therefore, we can estimate $A_2$ from the slope of the linear regression in Fig.~3(b), which gives $A_2 = 6.4(2)\unit{\mu{}m\cdot s^{1/3}}$.
Using this, we define the following, refined rescaled height:
\begin{equation}
    q'(x,t) \equiv \frac{h(x,t) - v_\infty t - A_2 t^{-1/3}}{(\Gamma t)^{1/3}} \simeq \chi + \mathcal{O}(t^{-1/3}). \label{eq:qprime}
\end{equation}
The difference between its mean $\expct{q'}$ and that of the Baik-Rains distribution $\expct{\chi_0}=0$ is shown by green solid disks in \figref{fig-cum}(a) and (b), which turns out to differ considerably from that of the usual rescaled height $\expct{q}$ (blue open circles).
Remarkably, with this refined rescaled height, the difference is found to decay as $t^{-1/3}$ over sufficiently long times [\figref{fig-cum}(b)].
Although the data at latest times seem to deviate slightly from this power law, we consider that this is probably due to small errors in the estimates of the rescaling parameters, in particular that of $\Gamma$, for which the value from the flat experiments was used.
Under this expectation, we can conclude that our experimental data clearly show convergence to the Baik-Rains distribution, at least up to the fourth-order cumulants.

\section{III. Two-Time Correlation}

\begin{figure}[t]  
 \centering
 \includegraphics[clip,width=0.8\hsize]{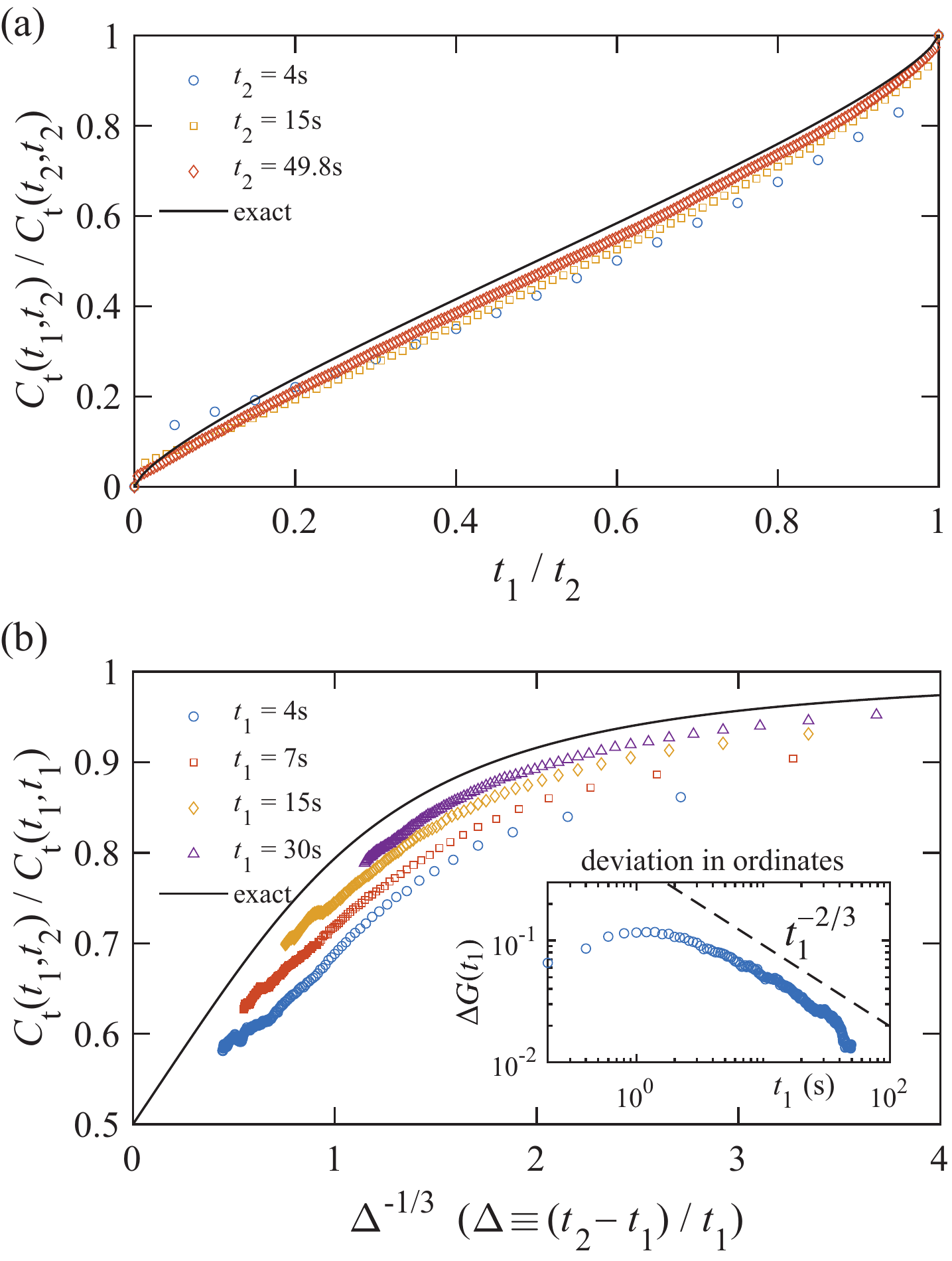}
 \caption{
Results on the two-time correlation for the Brownian interfaces.
(a) The two-time function normalized by the variance at time $t_2$, $F_\mathrm{t}(t_1,t_2)=C_\mathrm{t}(t_1,t_2)/C_\mathrm{t}(t_2,t_2)$, for different $t_2$ (legend), shown against $t_1/t_2$.
The same data as Fig.~4(c) in the main article are used, with more data points shown.
The black curve indicates Ferrari and Spohn's exact solution \cite{Ferrari.Spohn-SIG2016} $F_\mathrm{t}^\text{FBM}(t_1/t_2)$.
(b) The two-time function normalized by the variance at time $t_1$, $G_\mathrm{t}(t_1,t_2)=C_\mathrm{t}(t_1,t_2)/C_\mathrm{t}(t_1,t_1)$, for different $t_1$ (legend), shown against $\Delta^{-1/3}$ with $\Delta \equiv (t_2-t_1)/t_1$.
The black curve indicates Ferrari and Spohn's exact solution \cite{Ferrari.Spohn-SIG2016} $G_\mathrm{t}^\text{FBM}(\Delta^{-1/3})$.
The inset shows the deviation of the ordinates of the experimental data from the exact solution $G_\mathrm{t}^\text{FBM}(\Delta)$, as a function of $t_1$.
The dashed line is a guide for eyes indicating a power law $t_1^{-2/3}$.
}
 \label{fig-twotime}
\end{figure}

In this section we test agreement with Ferrari and Spohn's exact solution \cite{Ferrari.Spohn-SIG2016} on the two-time correlation function, defined by $C_\mathrm{t}(t_1,t_2)\equiv\expct{\delta{}h(x,t_1)\delta{}h(x,t_2)}$ with $\delta{}h(x,t)\equiv{}h(x,t)-\expct{h(x,t)}$.
Ferrari and Spohn's exact solution reads, with $\tau\equiv t_1/t_2$ (hence $0 \leq \tau \leq 1$) and in the limit $t_1,t_2 \to \infty$,
\begin{equation}
\frac{C_\mathrm{t}(t_1,t_2)}{C_\mathrm{t}(t_2,t_2)} \to F_\mathrm{t}^\text{FBM}(\tau) \equiv \frac{1}{2}\[1+\tau^{2/3}-(1-\tau)^{2/3}\],  \label{eq:FS1}
\end{equation}
 which is identical to the two-time correlation function of the fractional Brownian motion with Hurst exponent $1/3$ (abbreviated to \FBM).
The experimental data from the Brownian experiments are found to be close to this exact solution, and approaching it with increasing $t_2$ [\figref{fig-twotime}(a)].

To test whether the data obtained at finite times converge to this exact solution or not, we first note that the asymptotic behavior of \eqref{eq:FS1} in the limit $\tau \to 1$ ($t_2 \to t_1$) has been known since long before \cite{Kallabis.Krug-EL1999}.
The nontrivial limit is therefore $\tau \to 0$, corresponding to $t_1 \ll t_2$.
To study this limit with $t_1,t_2$ kept large, it is more convenient to fix $t_1$ and vary $t_2$, and normalize $C_\mathrm{t}(t_1,t_2)$ by the variance at $t_1$.
We therefore define
\begin{equation}
    G_\mathrm{t}(t_1,t_2) \equiv \frac{C_\mathrm{t}(t_1,t_2)}{C_\mathrm{t}(t_1,t_1)}
\end{equation}
and, following the convention adopted in Ref.~\cite{DeNardis.etal-PRL2017}, $\Delta \equiv (t_2-t_1)/t_1 = \tfrac{1}{\tau}-1$ as a time variable.
In this notation, Ferrari and Spohn's solution reads:
\begin{align}
    \frac{C_\mathrm{t}(t_1,t_2)}{C_\mathrm{t}(t_1,t_1)} \to G_\mathrm{t}^\text{FBM}(\Delta) \equiv \frac{1}{2}\[(\Delta+1)^{2/3} + 1 - \Delta^{2/3} \].  \label{eq:FS2}
\end{align}
This tends to $G_\mathrm{t}^\text{FBM}(\Delta) \simeq \frac{1}{2} + \frac{1}{3}\Delta^{-1/3}$ for large $\Delta$.
Note that the correlation remains strictly positive in the limit $\Delta\to\infty$, the property called the persistent correlation in Ref.~\cite{DeNardis.etal-PRL2017}.
Figure~\ref{fig-twotime} shows the solution \pref{eq:FS2}, together with the experimental data, plotted against $\Delta^{-1/3}$.
We can see that the data approach the theoretical curve with increasing $t_1$.
Interestingly, the finite-time data can be approximated by simply translating the theoretical curve downward.
We therefore fit the data by $G_\mathrm{t}(t_1,t_2) = G_\mathrm{t}^\text{FBM}(\Delta) - \Delta G(t_1)$ and plot $\Delta G(t_1)$ in the inset of \figref{fig-twotime}.
The result suggests $\Delta G(t_1) \sim t_1^{-2/3}$ for large $t_1$, indicating that the experimental data on the two-time correlation indeed seem to converge to Ferrari and Spohn's exact solution.

\section{Supplementary Movie Captions}

\subsection{Movie S1}

A typical realization of a flat interface, separating the metastable DSM1 (gray) and growing DSM2 regions (black).
The movie is played five times as fast.
The frame size is $3820\unit{\mu{}m} \times 3300\unit{\mu{}m}$.

\subsection{Movie S2}

A typical realization of a Brownian interface, separating the metastable DSM1 (gray) and growing DSM2 regions (black).
The movie is played five times as fast.
The frame size is $2730\unit{\mu{}m} \times 3300\unit{\mu{}m}$.

\pagebreak

\bibliography{IwaTak}